\documentclass{article}
\usepackage[utf8]{inputenc}
\usepackage{arabtex}
\usepackage{textcomp}
\usepackage[numbers]{natbib}
\usepackage{mathtools}
\usepackage{float}
\title{Design and performance of a terahertz photoconductive antenna with nano-crossfinger structure}
\author{Jitao Zhang $^{\dag}$\\ ECE Department,The University of Arizona, Tucson, AZ,85721\\
$^{\dag}$ \textit{jitaozhang@email.arizona.edu}}

\date{%
    \today% thats the default I guess
    \\[2\baselineskip]% Space between date and abstract
    \normalfont\normalsize%
    \parbox{0.8\linewidth}{%
{\bfseries Abstract}: Improving terahertz(THz) radiation power and/or optics-to-THz efficiency of the photoconductive antenna(PCA) is widely recognized as one of the most attractive and challenging missions in THz community. In this work, the design of a THz PCA with nano-crossfinger structure in the active region is proposed. The THz radiation properties of this PCA was demonstrated by finite-difference-time-domain method based on full-wave model. As a comparison, the PCA with nano-finger structure that promises enhanced THz radiation than conventional PCA was also analyzed numerically. The results indicate that the nano-crossfinger PCA can radiate even higher THz field than the nano-finger PCA, primarily due to the enhanced bias field within the nano-structure.
 }
}
\usepackage{natbib}
\usepackage{graphicx}

\begin{document}

\maketitle

\section{Introduction}
The existing broadband terahertz(THz) pulse sources can be usually generated through either electronic or optical procedure\cite{lee09}. On the electronic side, electron accelerator can radiate intense and ultra-broadband THz pulse. However, the bulk facility makes it only available at several places in the world\cite{carr2002,wu2013}. On the optical side, optical rectification, photoconductive antenna(PCA) and photo-induced gas plasma have been demonstrated before\cite{reimann2007}. Among them, the PCA is one of the earliest and most commonly used THz pulse sources due to its simple configuration and promising performance\cite{auston1984}. However, the biggest obstacle of a common PCA is its inefficient radiation, whose averaging power is only in the order of several $\mu W$. This problem stems from the low optics-to-THz conversion efficiency, and is eventually determined by the low quantum efficiency. In other words, most of the photo-excited carriers (i.e. electrons and holes) in the active region of a PCA will be recombined before they can arrive at the PCA's electrodes so that have little contribution to the THz radiation. Many efforts have been made to address this low-conversion-efficiency problem\cite{brener1996,kim2005,mar2011,winnerl2012,jitao_bowtooth}. Very recently, the idea of nano-structure PCA has attracted much attention\cite{park2012,singh2013,tanoto2013,berry2013,Huangyi2014}. The specific design of nano-structures within the bare gap of a common PCA can
not only optimize the distribution of the incoming laser beam but also improve the collection efficiency of the electrodes. Thereby, the enhancement of THz power as large as $50$ times has been demonstrated successfully\cite{berry2013}. In this work, a PCA with nano-crossfinger structure has been proposed, and its performance was numerically predicated by finite-difference-time-domain(FDTD) method based on full-wave model\cite{jitao2014}. Compared with the PCA with nano-finger structure, this PCA can radiate even higher THz field, which makes it a promising candidate of THz pulse source.

\section{Design and simulation of nano-crossfinger PCA}

\begin{figure}[h!]
\centering
\includegraphics[width=.8\textwidth]{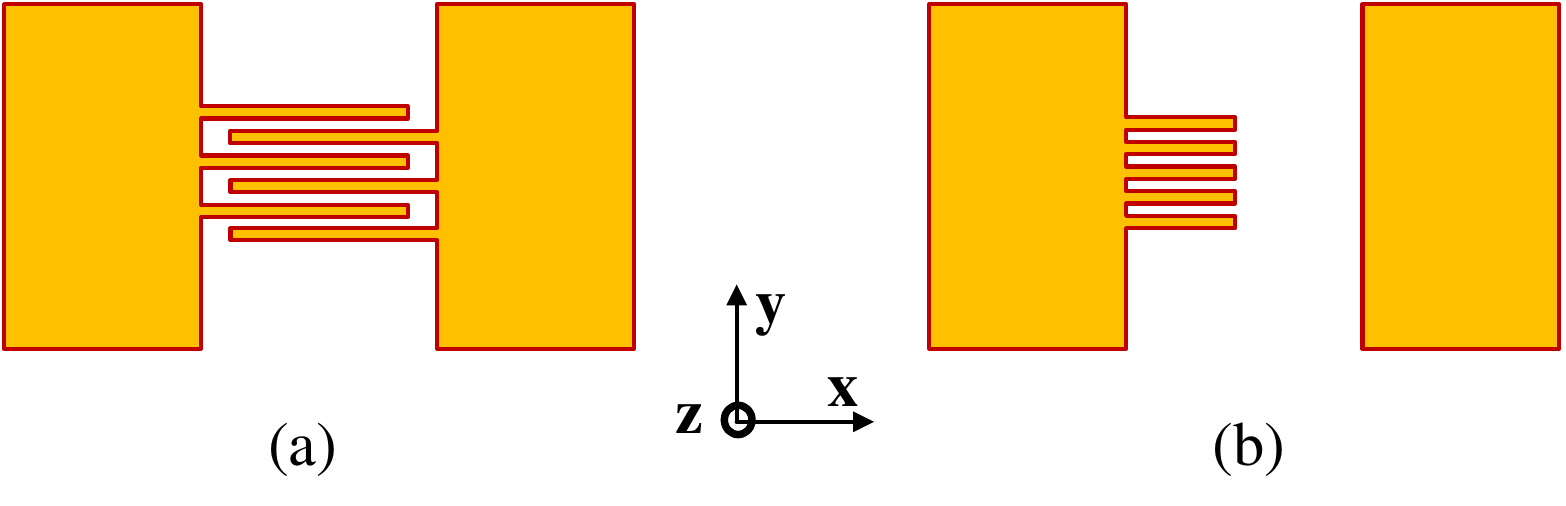}
\caption{Schematic structure of the (a) nano-crossfinger PCA and (b) nano-finger PCA. The left and right electrodes are anode and cathode, respectively. }
\label{fig1:structure}
\end{figure}

The structures of the nano-crossfinger and nano-finger PCAs are shown in Fig.\ref{fig1:structure}. Different from the nano-finger PCA shown in Fig.\ref{fig1:structure}(b), where all of the fingers contact the same electrode, the fingers of the nano-crossfinger PCA shown in Fig.\ref{fig1:structure}(a) are divided into half-and-half and contact with different electrodes in an interdigitating way. The fingers of both PCAs have the same width of $100 nm$ and separation of $100 nm$. The finger's length in Fig.\ref{fig1:structure}(a) and (b) is $1.7 \mu m$ and $1 \mu m$, respectively. Considering the available memory of the computer, the electrodes of both PCAs are $5 \mu m$ by $5 \mu m$ and have a gap size of $2 \mu m$ in between.

In our previous work\cite{jitao2014,jitao_para}, a full-wave simulation tool has been developed to predicate the radiation properties of the PCA based on FDTD method. The feature of this tool is that the multi-physical phenomena happening in the PCA, such as light-matter interaction, photo-excited carrier dynamics and full-wave propagation of the THz radiation, are considered and embodied in the simulation. Therefore, this tool can fulfill the needs of the comprehensive simulation of the PCA, in which almost all of the parameters that tightly related to the performance of the PCA can be involved, and the THz radiation can be predicated both in the near-field and far-field. This simulation tool was used in this work. The substrate has a dimension of $12 \mu m$ by $5 \mu m$ by $2.2 \mu m$ in $x$,$y$ and $z$ direction, respectively. The incident laser beam illuminated the nano-structure within the gap of the PCA. The distribution of optical absorption on the structure of fingers was simulated by COMSOL, and then was used as input data for FDTD simulation. All of the input parameters of the simulation are kept the same for both PCAs, and summarized in Table.1.

\begin{table}[H]
\caption{Parameters used in the simulation}
\centering
    \begin{tabular}{ll}
    \hline
    \textbf{Parameters}                        & \textbf{Values}                   \\ \hline
    Material                          & LT-GaAs                  \\
    %Doping density ($cm^{-3}$)       & 1.5E15                   \\
    Carrier lifetime ($ps$)             & electron: 0.1, hole: 0.4 \\
    Mobility ($cm^{2}/V\cdot s$) & electron: 200, hole: 30  \\
    Permittivity                      & 12.9                     \\
    Intrinsic concentration ($cm^{-3}$) & 2.1E6                      \\
    Absorption coefficient ($cm^{-1}$) & 1E4                     \\
    Laser wavelength ($nm$)             & 800                      \\
    Beam size ($\mu m$)       & $1$                  \\
    Averaging power ($mW$)       & $6$                  \\
    Pulse duration ($fs$)               & 80                       \\
%    Intensity ($W/cm^{2}$)           & 1E9                      \\
    DC voltage ($V$)           & 0.5                      \\ \hline
    \end{tabular}
\label{inputdata}
\end{table}

\section{Simulation results}

\begin{figure}[H]
\centering
\includegraphics[width=1\textwidth]{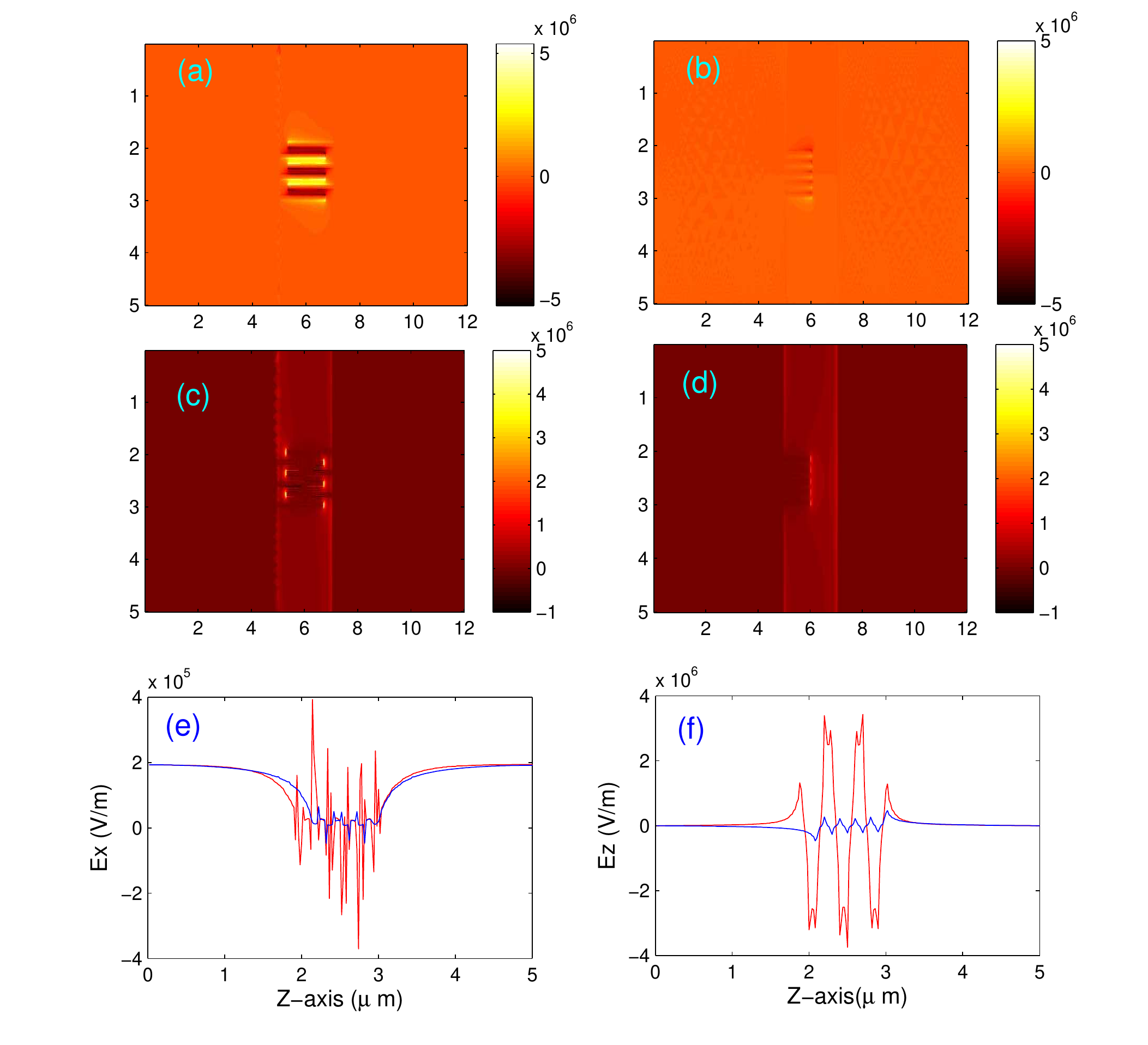}
\caption{Distribution of the DC bias field. (a) and (c) are $E_{z}$ and $E_{x}$ on the top surface of nano-crossfinger PCA, respectively. (b) and (d) are $E_{z}$ and $E_{x}$ on the top surface of nano-finger PCA, respectively. The horizontal and vertical axises of (a) to (d) represent $x$ and $z$ directions with unit of $\mu m$, respectively. The colorbars have unit of $V/m$. (e) and (f) are cut lines along $z$ axis at the center of $x$ axis. The red and blue solid lines represent the field of nano-crossfinger and nano-finger PCAs, respectively.}
\label{fig:dc}
\end{figure}

Figure \ref{fig:dc} shows the distribution of the DC bias field of both PCAs. Fig.\ref{fig:dc}(a) to (d) are $E$ fields on the top surface($XOY plane$) of two PCAs. It indicates that $E_{z}$ field of the nano-crossfinger PCA is greatly enhanced, and $E_{x}$ field is enhanced as well. The cut lines drawn in Fig.\ref{fig:dc}(e) and (f) show the enhancement much clearly. The $E_{z}$ field of the nano-crossfinger PCA is approximately $10$ times of that of the nano-finger PCA at the peak. Since the bias field is the source of the force that drives the photo-excited carriers to flow to the electrodes, enhanced local bias field will promise larger photo-current and thus higher THz radiation power. One may find in Fig.\ref{fig:dc}(e) that $E_{x}$ is even larger at the bare region of the gap. However, different from the case of the region having finger structure, most of the photo-excited carriers in bare region will have much less contribution to THz radiation since the transport path is too long to be reachable before recombination.

\begin{figure}[H]
\centering
\includegraphics[width=1\textwidth]{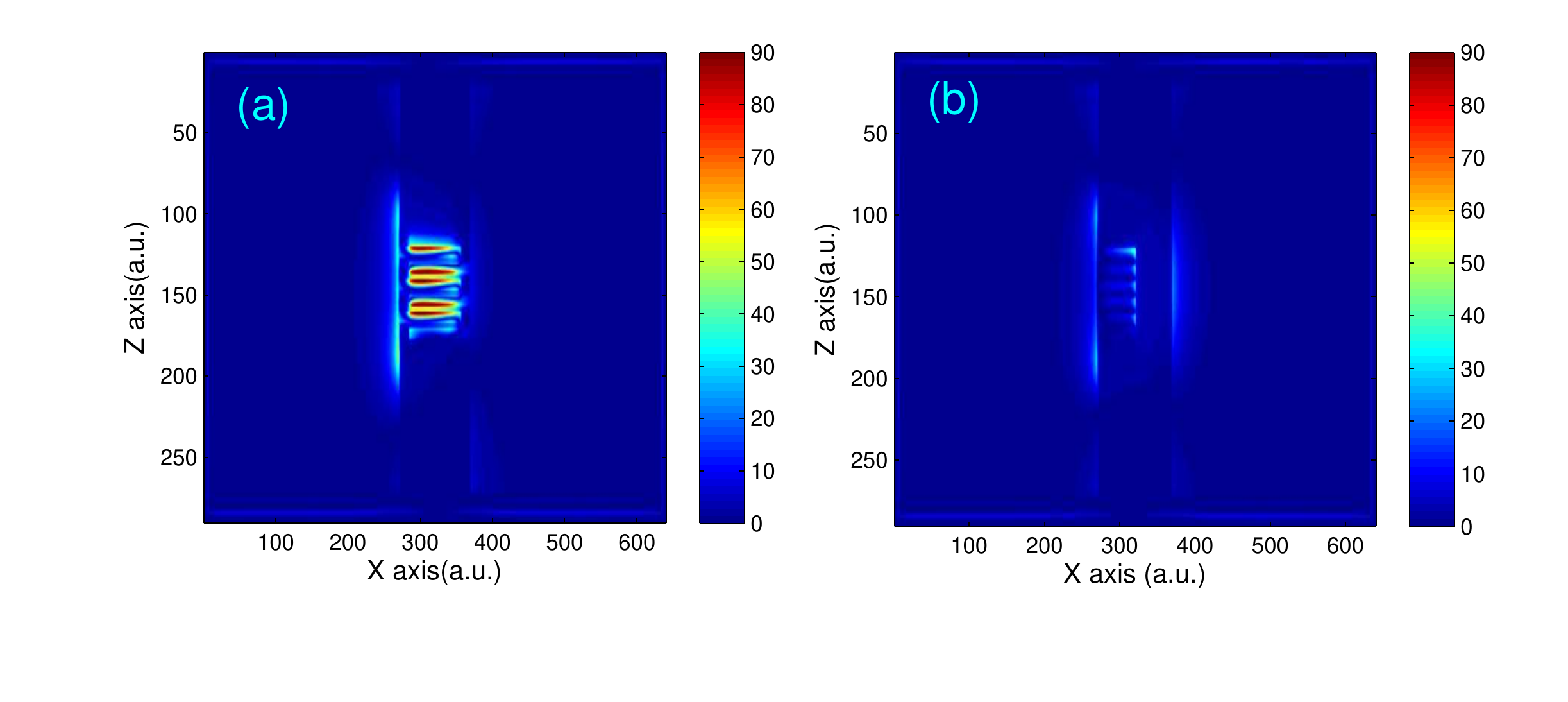}
\caption{Transient E field of (a) nano-crossfinger PCA and (b) nano-finger PCA at $0.15ps$ on the top surface of the PCAs. The $x$ and $z$ aixis are counted by mesh number. The colorbars have unit of $V/m$.}
\label{fig:E_total}
\end{figure}

\begin{figure}[H]
\centering
\includegraphics[width=1\textwidth]{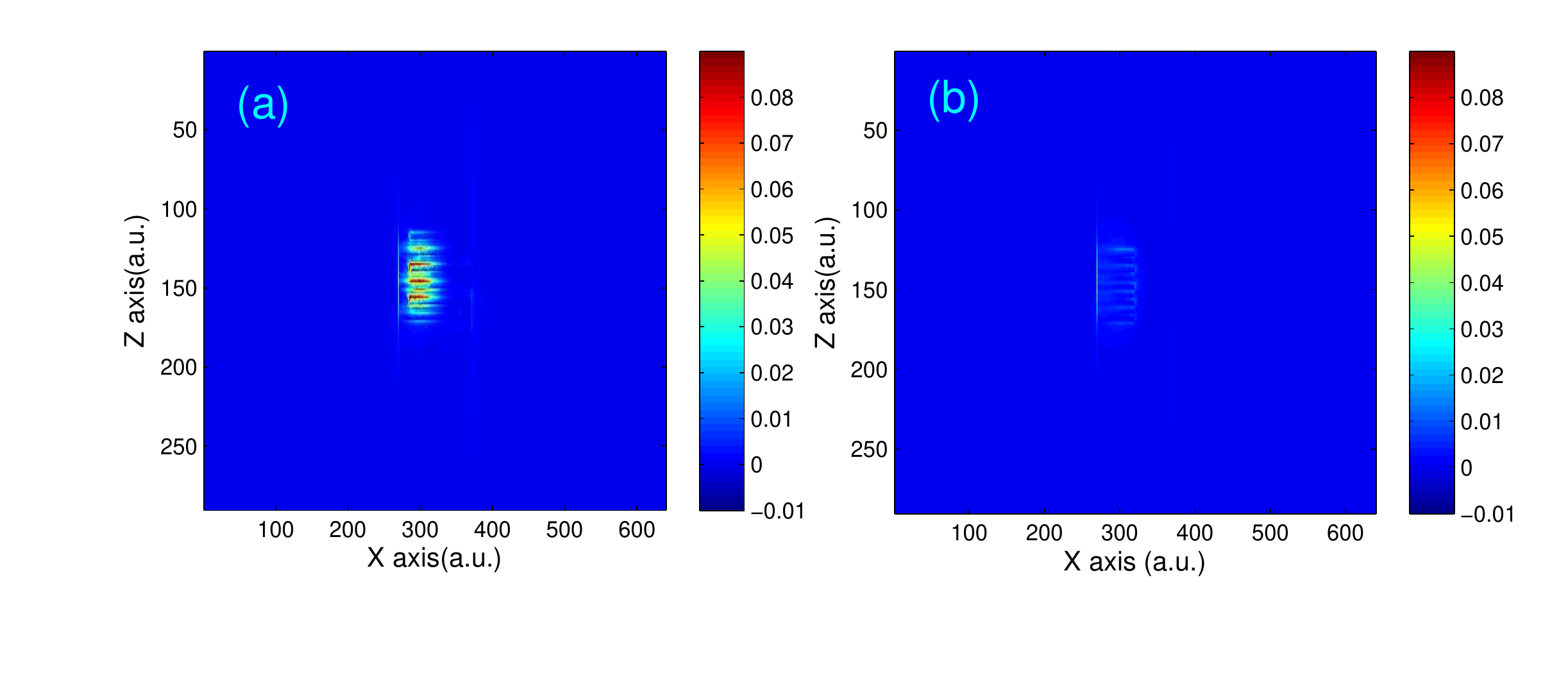}
\caption{Transient current density of (a) nano-crossfinger PCA and (b) nano-finger PCA at $0.15ps$ on the top surface of the PCAs. The $x$ and $z$ aixis are counted by mesh number. The colorbars have unit of $A/\mu m^{2}$.}
\label{fig:J_total}
\end{figure}

\begin{figure}[H]
\centering
\includegraphics[width=0.7\textwidth]{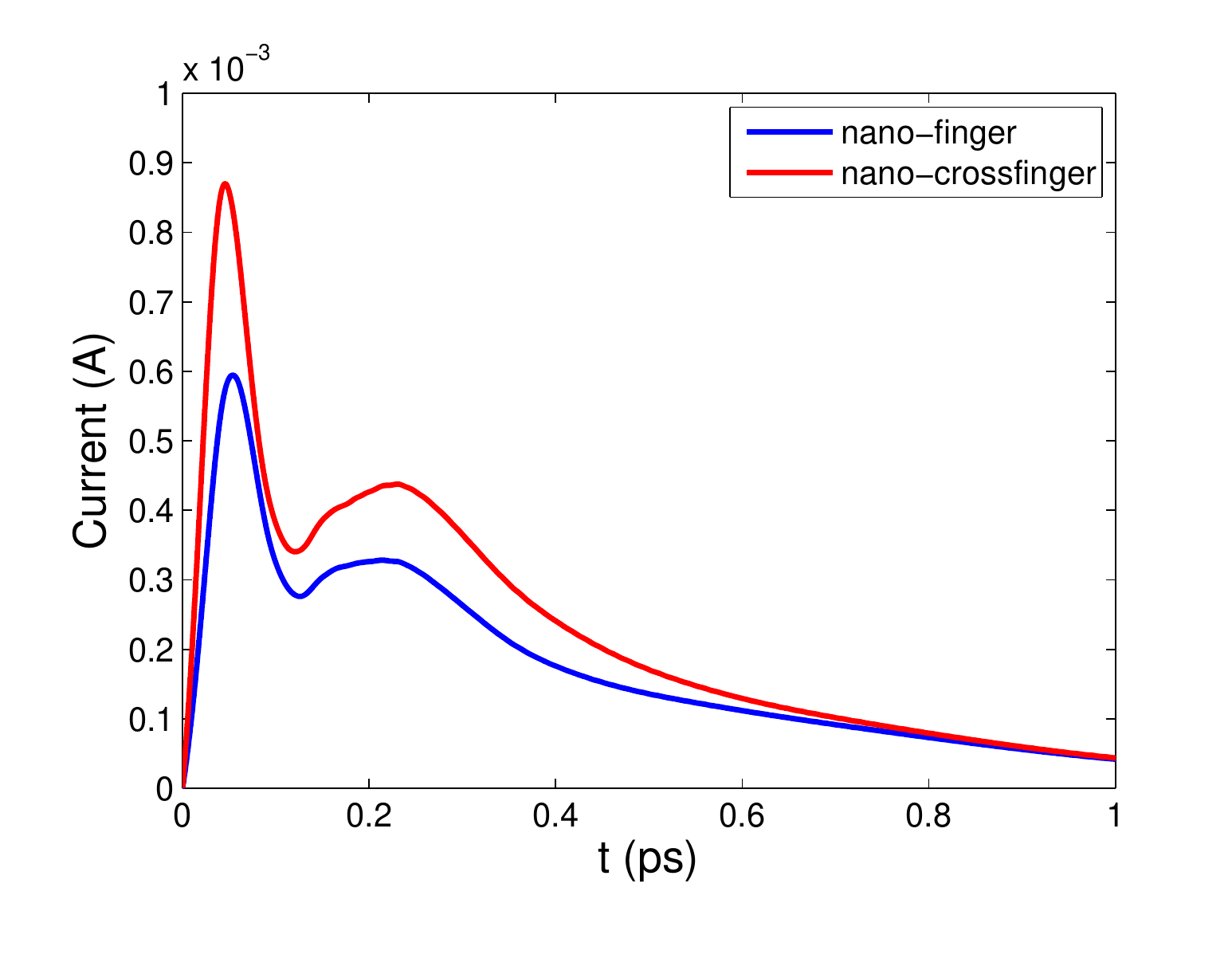}
\caption{Time-domain photo-excited current of the cross-section that parallels $XOY$ plane and passes through the center of the laser beam.}
\label{fig:current}
\end{figure}

The FDTD simulation was run for $1 ps$ with time step of $0.033fs$ and mesh size of $20 nm$, and the peak of the laser pulse was introduced at $0.15 ps$. The transient $E$ field at $0.15 ps$ on the top surface of the PCAs is shown in Fig.\ref{fig:E_total}, which clearly indicates the enhancement of the nano-crossfinger PCA. In addition, Figure \ref{fig:J_total} shows the transient current density on the top surface of the PCAs. It confirmed the prediction above about the 
function of the bias field. The photo-excited current within the gap region of the nano-crossfinger PCA is larger than that of the nano-finger PCA. This is further confirmed by the photo-excited current of the cross-section that parallels $XOY$ plane and passes through the center of the laser beam, as shown in Fig.\ref{fig:current}.

\begin{figure}[H]
\centering
\includegraphics[width=.9\textwidth]{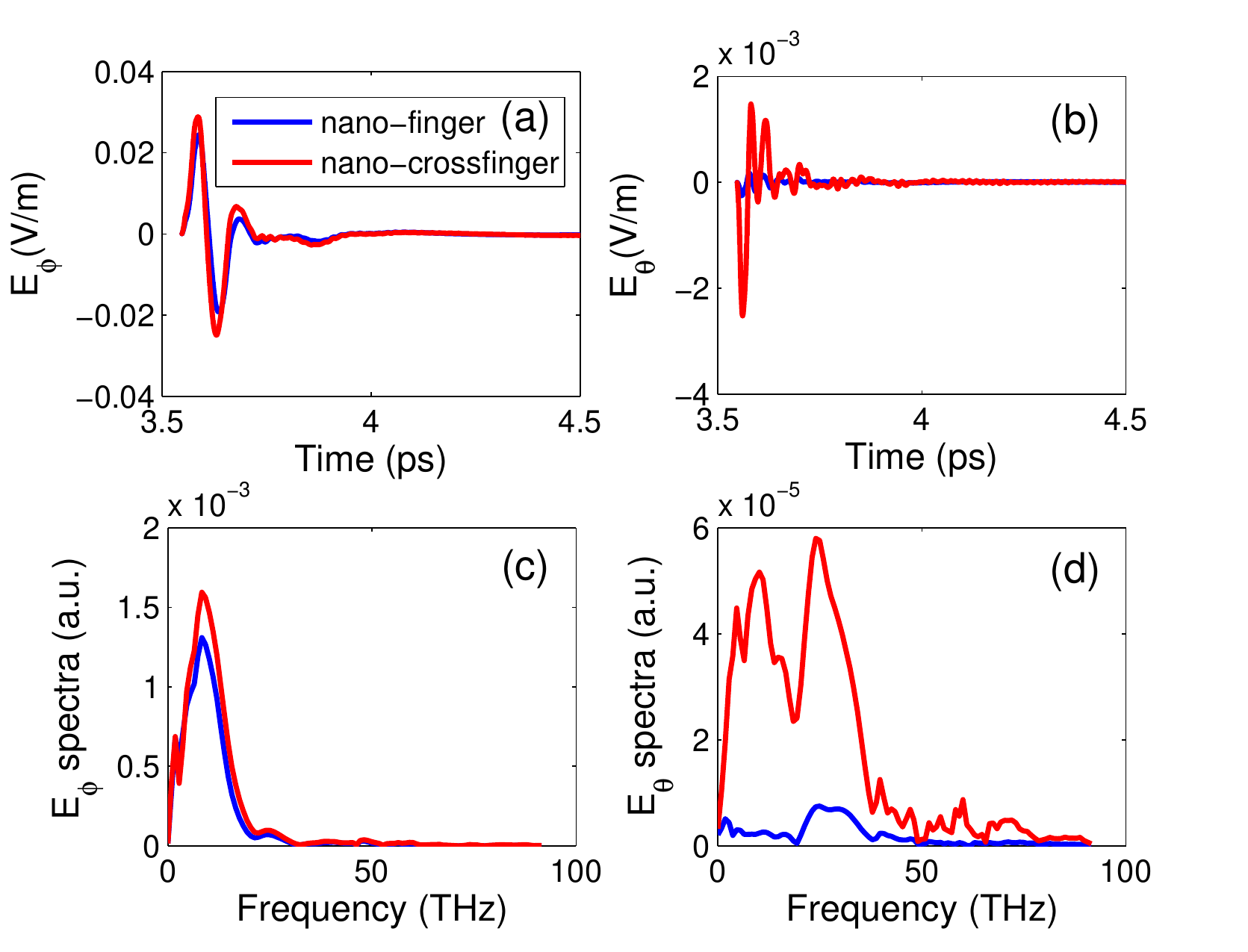}
\caption{Far-field THz radiation. (a) and (b) are time-domain E fields $E_{\phi}$ and $E_{\theta}$. (c) and (d) are corresponding spectra calculated by Fast-Fourier-Transform. The red and blue solid lines represent data of nano-crossfinger and nano-finger PCAs, respectively.} 
\label{fig:farfield}
\end{figure}

The far-field THz radiation was calculated at the point right below the PCA (along $z$ axis) with a distance of $200 mm$ away from the center of the PCA. Figure \ref{fig:farfield} shows far-field THz radiations of nano-crossfinger and nano-finger PCAs. The calculated polarized components $E_{\phi}$ and $E_{\theta}$ of the THz field are along $x$ and $y$ axises, respectively. It is shown that the nano-crossfinger PCA outperforms nano-finger PCA regarding the far-field THz radiation. For example, $E_{\phi}$ of nano-crossfinger PCA is $1.2$ times of that of nano-finger PCA at peak frequency. Moreover, the enhancement of $E_{\theta}$ can be as large as $20$ times at $10 THz$! This is mainly due to the strongly enhanced bias field in $z$ direction, as shown in Fig.\ref{fig:dc}(f).

\section{Conclusion}

A nano-crossfinger PCA has been designed, and its performance was numerically analyzed and compared with that of another nano-finger PCA. The enhanced THz radiation has been found for this new PCA, which is mainly due to the optimized distribution of the bias field. The proposed PCA is conceptually different from previous interdigitated PCA design. In the interdigitated PCA, the structures have micrometer dimensions, and the whole PCA actually works as an array of common coplanar stripline PCAs. In nano-crossfinger PCA, however, the typical dimension of the fingers is in nanometer. As such, the fingers will not only optimize the distribution of the incoming laser beam by diffraction, but also shorten the transport path of the photo-excited carriers. Therefore, the optics-to-THz efficiency as well as the THz radiation field can be strongly enhanced. Regarding the future work, the far-field pattern and the averaging outpower of the nano-crossfinger PCA will be evaluated numerically, and the experimental verification is also under consideration.

\section*{Acknowledgements}
This work can not be done without the contribution of the following colleagues. They are Mingguang Tuo, Min Liang and Hao Xin. They will be listed as co-authors when we consider publishing this work in a journal.

\bibliographystyle{plain}
\bibliography{references}
\end{document}